\documentclass[
    ,final            
  ]
  {aipproc}

\layoutstyle{6x9}

\begin{document}

\title{The crust of neutron stars}



\author{N. Chamel}{
  address={Institut d'Astronomie et d'Astrophysique, Universit\'e Libre de Bruxelles, CP226, Boulevard du Triomphe, 1050 Brussels (Belgium)}
}

\begin{abstract}

The structure of the crust of a neutron star is completely determined by the experimentally measured nuclear masses up to a density of the order of
$10^{11}$ g.cm$^{-3}$. At higher densities, the composition of the crust still remains uncertain, mainly due to the presence of ``free'' superfluid neutrons which
affect the properties of the nuclear ``clusters''. After briefly reviewing calculations of the equilibrium structure of the crust, we point out that
the current approach based on the Wigner-Seitz approximation does not properly describe the unbound neutrons. We have recently abandoned
this approximation by applying the band theory of solids. We have shown that the dynamical properties of the free neutrons
are strongly affected by the clusters by performing 3D calculations with Bloch boundary conditions.
\end{abstract}

\maketitle


\section{Introduction}

At the end point of the stellar evolution, neutron  stars are the compact remnants of core collapse supernova explosions. Born with temperatures as high
as $10^{11}-10^{12}$ K, the star rapidly cools down by emitting neutrinos and photons. A few hours after its birth, the temperature of the star falls below
about $10^9$ K and the external layers crystallize into a solid crust (for a general review of neutron star crust, see \cite{pethickravenhall95, haensel01b}).
Within hundreds of years, the interior of the star becomes isothermal with temperatures typically less than $10^6$ K ($\sim 0.1$ keV).

From the nuclear physics point of view, a neutron star is a huge nucleus containing about $A\sim 10^{57}$ baryons and a proton fraction of the order of
10\%. A rough estimate of the radius and the mass of the star from the liquid drop model yields $R=r_0 A^{1/3} \sim 10$ km and $M=A m_p\sim M_\odot$
 respectively. The solid crust surrounding the star plays more or less the role of the neutron skin in heavy nuclei. Indeed correlations have been
 established between the neutron skin in lead $^{208}$Pb and the density at which the crust melts into a uniform liquid \cite{Horowitz01}. This gross picture however should not be taken too far since the conditions prevailing inside neutron stars
are very different from those inside isolated nuclei. 

The crust of a neutron star represents only a few percent of the mass of the star but plays a crucial role
for its evolution (magnetic field, cooling, bursts, starquakes, spinning-down, glitches, free precession, non axial deformations giving rise to the emission of gravitational waves). We shall
briefly review calculations of the equilibrium structure of the crust in the first section. The next section will be devoted to the neutron superfluidity in the crust. In the last section, we will show how the description of the inner crust can be improved by applying the band theory of solids and we will discuss some recent results.

\section{Structure and composition of the crust}
\label{structure}

In the following it will be assumed that the complete thermodynamical equilibrium at zero temperature with respect
to all interactions has been reached and therefore the matter is in its lowest energy state (this excludes 
newly-born hot neutron stars and neutron stars accreting matter from a companion star which will not be discussed here). This assumption is usually known as the cold catalyzed matter hypothesis. The ground state of the crust is obtained by minimizing the total energy per nucleon under the assumption of beta equilibrium and electroneutroneutrality. The crust is further supposed to be formed of a perfect crystal with a single nuclear species at lattice sites (for a discussion of possible deviations from this idealized model, see for instance \cite{Haensel06} and references therein).

\subsection{outer crust}

At densities below $\sim 10^7$ g.cm$^{-3}$, the ground state of matter is a mixture of electrons and iron $^{56}$Fe (the atoms are fully ionized
at densities above $\sim 10^4$ g.cm$^{-3}$). At higher densities,
nuclei become increasingly neutron rich due to inverse beta decay. Following the classical paper of Baym, Pethick and Sutherland  \cite{BPS}, the total energy density in a given layer
can be written as
\begin{equation}
\label{equation:total_energy}
\varepsilon_{\rm tot} = n_N E\{A,Z\} +\varepsilon_e+\varepsilon_L
\end{equation}
where $n_N$ is the number density of nuclei, $E\{A,Z\}$ is the energy of a nucleus with $Z$ protons and $A-Z$ neutrons, $\varepsilon_e$ is the electron energy density
and $\varepsilon_L$ is the lattice energy density. At densities $\rho\gg 10^6$ g.cm$^{-3}$ the electrons can be described as a relativistic Fermi gas.
Assuming point like nuclei (the lattice spacing being very large compared to the size of the nuclei),
the lattice energy density can be expressed as

\begin{equation}
\label{equation:lattice_energy}
\varepsilon_L = - c\left(\frac{4\pi}{3}\right)^{1/3} Z^2 e^2 n_e^{4/3}  \, ,
\end{equation}
where $c$ is a coefficient which depends on the lattice structure. For cubic structures, this coefficient is respectively equal to
$0.89593$, $0.89588$ and $0.88006$ for body centered, face centered and simple cubic lattice which suggests that the crust
 crystallizes in a body centered cubic lattice.

The main physical input is therefore the energy of a nucleus. The structure of the outer crust is completely determined by the experimental
nuclear masses up to a density of the order $\rho \sim 6 \times 10^{10}$ g.cm$^{-3}$ \cite{Ruster06}. At higher densities the nuclei are so neutron rich that
the energy $E\{A,Z\}$ must be extrapolated. The composition of the nuclei in these layers is thus model dependent.
Nevertheless most models predict the existence of nuclei with the magic neutron numbers $N=50, 82$, thus revealing the crucial role played by shell effects.
The nuclei present at the bottom of the outer crust may be experimentally studied
 in the near future by several facilities, such as for instance FAIR at GSI \footnote{\url{http://www.gsi.de/fair/index_e.html}}, ISAC at TRIUMPH
 \footnote{\url{http://www.triumf.info/public/about/isac.php}}, SPIRAL 2 at GANIL \footnote{\url{http://www.ganil.fr/research/developments/spiral2/index.html}} and by the RIA project \footnote{\url{http://www.phy.anl.gov/ria/}}. The structure of
 the outer crust is shown in table \ref{table:BSk8_outer_crust} for one particular representative recent model.

\begin{table}
\begin{tabular}{|c c c c c c c|}
\hline
$\mu$ [MeV] & $\mu_e$ [MeV] & $\rho_\mathrm{max}$ [g/cm$^3$] & $n_b$ [cm$^{-3}$] & Element & $Z$ & $N$ \\
\hline
\hline
930.60 &  0.95 & $8.02 \times 10^{6}$  &  $4.83 \times 10^{30}$ &  ${}^{56}$Fe & 26 & 30 \\
931.32 &  2.61 & $2.71 \times 10^{8}$  &  $1.63 \times 10^{32}$ &  ${}^{62}$Ni & 28 & 34 \\
932.04 &  4.34 & $1.33 \times 10^{9}$  &  $8.03 \times 10^{32}$ &  ${}^{64}$Ni & 28 & 36 \\
932.09 &  4.46 & $1.50 \times 10^{9}$  &  $9.04 \times 10^{32}$ &  ${}^{66}$Ni & 28 & 38 \\
932.56 &  5.64 & $3.09 \times 10^{9}$  &  $1.86 \times 10^{33}$ &  ${}^{86}$Kr & 36 & 50 \\
933.62 &  8.38 & $1.06 \times 10^{10}$ &  $6.37 \times 10^{33}$ &  ${}^{84}$Se & 34 & 50 \\
934.75 & 11.43 & $2.79 \times 10^{10}$ &  $1.68 \times 10^{34}$ &  ${}^{82}$Ge & 32 & 50 \\
935.89 & 14.61 & $6.07 \times 10^{10}$ &  $3.65 \times 10^{34}$ &  ${}^{80}$Zn & 30 & 50 \\
\hline
936.44 & 16.17 & $8.46 \times 10^{10}$ &  $5.08 \times 10^{34}$ &  ${}^{82}$Zn & 30 & 52 \\
936.63 & 16.81 & $9.67 \times 10^{10}$ &  $5.80 \times 10^{34}$ & ${}^{128}$Pd & 46 & 82 \\
937.41 & 19.16 & $1.47 \times 10^{11}$ &  $8.84 \times 10^{34}$ & ${}^{126}$Ru & 44 & 82 \\
938.12 & 21.35 & $2.11 \times 10^{11}$ &  $1.26 \times 10^{35}$ & ${}^{124}$Mo & 42 & 82 \\
938.78 & 23.47 & $2.89 \times 10^{11}$ &  $1.73 \times 10^{35}$ & ${}^{122}$Zr & 40 & 82 \\
939.47 & 25.77 & $3.97 \times 10^{11}$ &  $2.38 \times 10^{35}$ & ${}^{120}$Sr & 38 & 82 \\
939.57 & 26.09 & $4.27 \times 10^{11}$ &  $2.56 \times 10^{35}$ & ${}^{118}$Kr & 36 & 82 \\
\hline
\end{tabular}
\caption{Sequence of nuclei in the outer crust of
non-accreting cold neutron stars calculated by R\"uster {\it et al.} \cite{Ruster06}
(a theoretical nuclear mass table was used for the lower part). 
The last line corresponds to the neutron drip point.}
\label{table:BSk8_outer_crust}
\end{table}

\subsection{inner crust}

At the bottom of the outer crust, the nuclei become so neutron rich that some neutrons are no longer bound. The nuclear lattice then coexists with
a neutron gas. The transition occurs when the electron Fermi energy becomes comparable to the
binding energy of the protons in the nuclei. At the drip, the neutron excess $\delta=(1-2A/Z)$ can be estimated as \cite{Haensel06}
\begin{equation}
\label{equation:drip_delta}
\delta_{\rm drip} = \sqrt{1-E_0/S_0}-1 \, ,
\end{equation}
where $E_0$ is the energy per nucleon of infinite symmetric nuclear matter and $S_0$ the symmetry energy.
To lowest order in $\delta$, the density threshold for the onset of neutron drip is approximately given by
\begin{equation}
\label{equation:drip_density}
\rho_{\rm drip} \simeq (S_0 \delta_{\rm drip})^3 \times  10^9  \, {\rm g}.{\rm cm}^{-3} \, .
\end{equation}
For $E_0=-16$ MeV and $S_0=32$ MeV, we find $\delta_{\rm drip}\simeq 0.225$ and $\rho_{\rm drip} \simeq 4\times 10^{11}$ g.cm$^{-3}$
which is in remarkable agreement with the value obtained from more realistic nuclear models. The simple
estimates (\ref{equation:drip_delta}) and (\ref{equation:drip_density}) illustrate the importance of
the symmetry energy on the physics of neutron stars (see also \cite{Steiner05} and references therein). It has been 
very recently shown using a liquid drop model that the composition of the inner crust is very sensitive to the density dependence of the symmetry energy \cite{Oyamatsu06}. 

The inner crust of a neutron star is a unique environment which is not accessible in the laboratory due to the presence of the ``free'' neutron gas.
In the following we shall thus refer to the ``nuclei'' in the inner crust as ``clusters'' in order to emphasize these peculiarities. The structure of the inner
crust has been studied using various approaches, mainly liquid drop and semi-classical models (for a recent review, see for instance \cite{Haensel06}).

The most realistic calculations of the structure of the inner crust were pioneered by the work of Negele\&Vautherin \cite{NV73}. 
Until very recently\cite{Baldo06b}, it was the only fully self-consistent quantum calculation. 
Expanding the density matrix in relative and c.m. coordinates,
Negele\&Vautherin derived a set of effective equations for the nucleons, which closely resemble those obtained in the
Hartree-Fock approximation with Skyrme forces. They determined the structure of the inner crust by minimizing the total energy
per nucleon in the Wigner-Seitz sphere, treating the electrons as a relativistic Fermi gas.
As pertains the choice of boundary conditions, they imposed that wavefunctions with even parity (even $\ell$)
and the radial derivatives of wavefunctions with odd parity (odd $\ell$) vanish on the sphere
$r=R_{\rm cell}$. This prescription yielded roughly uniform density outside the nuclear clusters. The remaining spurious
fluctuations were removed at each iteration step by averaging the densities in the vicinity of the cell edge.

\begin{figure}
  \includegraphics[height=.2\textheight]{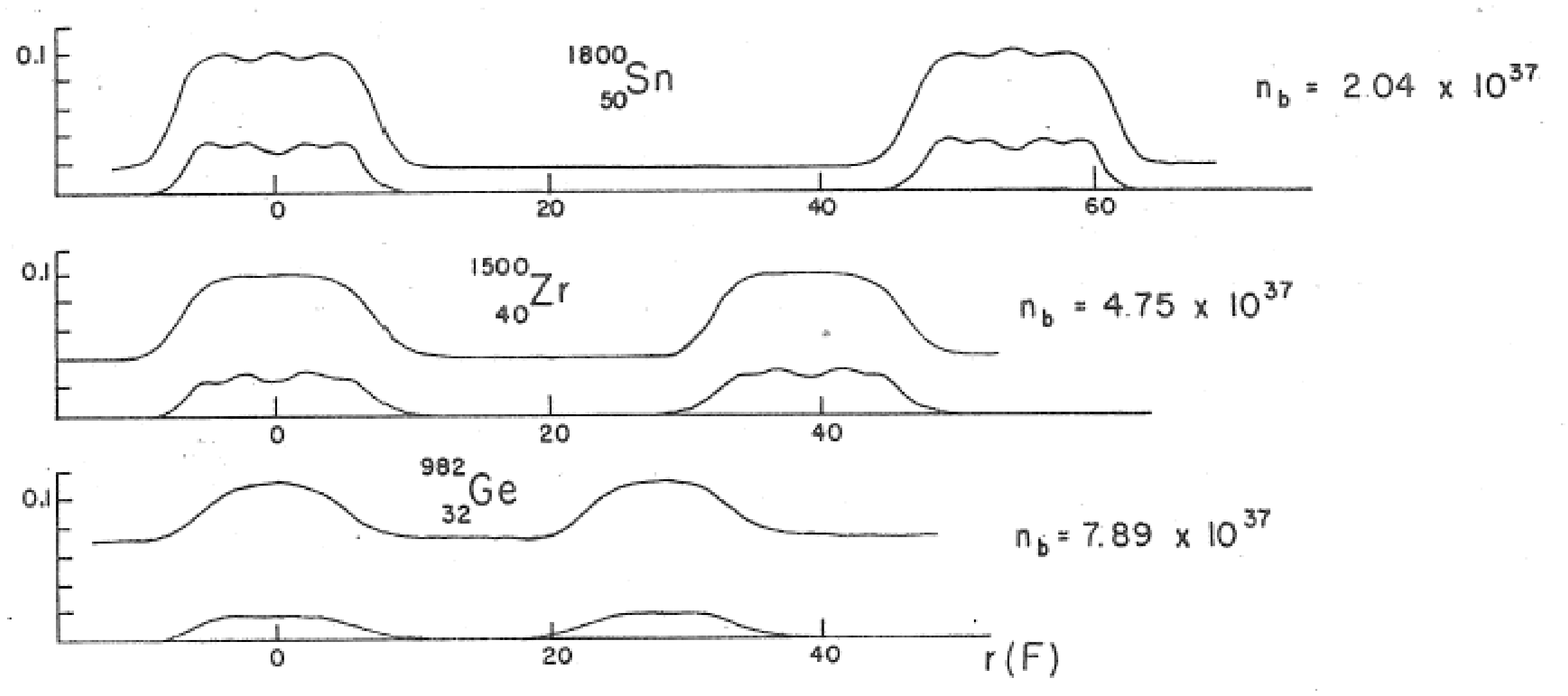}
  \caption{Nucleon number densities (in fm$^{-3}$) along the axis joining two adjacent Wigner-Seitz cells of the inner crust of neutron stars
   for a few  baryon densities $n_b$ (in cm$^{-3}$) as calculated by Negele\&Vautherin \cite{NV73}.}
  \label{fig1}
\end{figure}

The composition of the crust is shown on table \ref{table:NV_inner_crust}. These results are qualitatively similar to those obtained with liquid drop
and semiclassical models. The remarkable distinctive feature is the existence of strong proton shell effects
with a predominance of nuclear clusters with $Z=40$ and $Z=50$. Neutron shell effects are also important (while not obvious from the table)
as can be inferred from the density fluctuations inside the clusters on figure \ref{fig1}. This figure also shows that shell effects disappear
at high densities where the matter becomes nearly homogeneous.

\begin{table}
\begin{tabular}{|c c c c c c|}
\hline
$\mu_n$ [MeV] & $\mu_p$ [MeV]  & $n_b$ [cm$^{-3}$] & Element & $Z$ & $N$ \\
\hline
\hline
0.2 &  -26.8 & $2.79 \times 10^{35}$ & ${}^{180}$Zr & 40 & 140 \\
0.3 &  -29.4 & $4 \times 10^{35}$ & ${}^{200}$Zr & 40 & 160 \\
0.6 &  -29.5 & $6 \times 10^{35}$ & ${}^{250}$Zr & 40 & 210 \\
1.0 &  -28.5 & $8.79 \times 10^{35}$ & ${}^{320}$Zr & 40 & 280 \\
1.4 &  -29.4 & $1.59 \times 10^{36}$ & ${}^{500}$Zr & 40 & 460 \\
2.6 &  -33.6 & $3.73 \times 10^{36}$ & ${}^{950}$Sn & 50 & 900 \\
3.3 &  -34.5 & $5.77 \times 10^{36}$ & ${}^{1100}$Sn & 50 & 1050 \\
4.2 &  -35.8 & $8.91 \times 10^{36}$ & ${}^{1350}$Sn & 50 & 1300 \\
6.5 &  -43.6 & $2.04 \times 10^{36}$ & ${}^{1800}$Sn & 50 & 1750 \\
10.9 &  -54.0 & $4.75 \times 10^{37}$ & ${}^{1500}$Zr & 40 & 1460 \\
15 &  -68.3 & $7.89 \times 10^{37}$ & ${}^{980}$Ge & 32 & 950 \\
\hline
\end{tabular}
\caption{Sequence of nuclei in the inner crust of
non-accreting cold neutron stars calculated by Negele \& Vautherin \cite{NV73}.
$N$ is the total number of neutrons in the W-S sphere.}
\label{table:NV_inner_crust}
\end{table}

\subsection{``pasta'' phases}

The equilibrium composition of the clusters is the result of the competition between Coulomb and surface energies.
At low densities, the lattice energy (\ref{equation:lattice_energy}) is a small contribution to the total Coulomb
energy and nuclei are spherical. However at the bottom of the crust, the size of the cluster is of the same order
of magnitude as the lattice spacing and consequently the lattice energy represents a large reduction of the total Coulomb energy
(this reduction is about 15 \% at the neutron drip). This means that the nuclear clusters may be strongly deformed in the
high density layers of the inner crust. Reasoning by analogy with percolating networks, Ogasawara\&Sato\cite{Ogasawara82}
suggested that a transion to an ``infinite network of linked nuclei'' might occur at the bottom of the crust. The possibility of
non spherical nuclear clusters, referred as ``pastas'', was considered by
Ravenhall {\it et al.} \cite{Ravenhall83} and Hashimoto \&Yamada \cite{Hashimoto84} who found from compressible liquid drop models,
that as the density increases, the nuclei merge into cylinders (``spaghetti'')  followed by slabs (``lasagna''), cylindrical tubes and bubbles (``swiss cheese'').
The pasta phases cover a small range of densities near the crust-core interface. Nevertheless they may represents up to half of the mass of the crust.
The existence of these phases may have important astrophysical consequences for the gravitational wave emission and for pulsar glitches
by changing the elastic properties of the crust\cite{Pethick98}, for the cooling of neutron stars by allowing direct URCA processes \cite{Gusakov04} and
enhancing the heat capacity \cite{Blasio95,Elgaroy96} and for core-collapse supernovae \cite{Horowitz05}. These pasta phases have been studied by various nuclear models,
from liquid drop calculations to quantum molecular dynamic simulations (for a recent review see for instance \cite{Watanabe05} and references therein).
However a few models do not predict the existence of such phases \cite{Lorenz93, Cheng97, Douchin00,Maruyama05}. The energy differences between the various shapes are very small,
typically less than $\sim$ keV.fm$^{-3}$, and as a result the structure of the crust is very sensitive to small differences between nuclear models.

\section{Neutron superfluidity in the crust}
\label{superfluidity}

The possibility of superfluidity in neutron stars was suggested a long time ago by Migdal \cite{Migdal59}, only two years after the formulation of the BCS theory
of electron superconductivity. Microscopic calculations performed in pure neutron matter indicate that at densities below saturation density,
neutrons are bound in Cooper pairs which condense into a $^1$S$_0$ superfluid phase. However the density range for superfluidity and the magnitude of the pairing gap
still remain uncertain due to different approximations of medium polarization and self-energy effects which tend to suppress the pairing as compared to mean field
calculations with bare nucleon-nucleon forces (for a review see for instance \cite{Dean03,Baldo05b}).

The situation is even more uncertain in the crust owing to the small nuclear asymmetry and to the presence of inhomogeneities (for a recent review, see
\cite{Baldo05b}). Superfluidity of the ``free'' neutrons induces superfluidity of the bound neutrons inside the clusters and {\it vice versa}. This proximity
effect tends to smooth the spatial variations of the pairing field (see \cite{Sandulescu04} and references therein).
The effects of neutron superfluidity on the equilibrium structure of the crust have been considered by Baldo
{\it et al.} \cite{Baldo05} in the density functional theory generalized to account for nucleon pairing. The nuclear lattice was treated in the Wigner-Seitz approximation.
They determined the ground state of the crust
at the baryon density $\rho\simeq 1.9\times 10^{13}$ g.cm$^{-3}$, for which the neutron pairing is expected to be the strongest. They found that
the structure of the crust is significantly affected by the pairing. Indeed, the charge of the cluster $Z\simeq 52$ and the radius of the Wigner-Seitz sphere
$R_{\rm cell}\simeq 32$ fm are increased by about $\delta Z \simeq 8$
and $\delta R_{\rm cell}\simeq 3.5$ fm respectively compared to calculations without including the pairing. However they emphasized that the results are very sensitive to the choice of the energy functional. These conclusions have been very recently confirmed by calculations at other densities \cite{Baldo06b}. This urge the need for a better understanding of pairing correlations in 
inhomogeneous neutron star crust matter.

\section{From nuclear to solid state physics}
\label{solid}

\subsection{Wigner-Seitz approximation}

In the quantum calculations briefly reviewed in the previous sections, the nuclear lattice was treated in the Wigner-Seitz approximation. This approximation
however does not properly takes into account the unbound neutrons which are artificially confined inside the Wigner-Seitz sphere. This approximation leads to
spurious fluctuations of the neutron densities and of the neutron pairing field \cite{Montani04}. As a result calculations of the equilibrium structure of the crust and of
crustal superfluidity are contaminated by unphysical shell effects which are very sensitive to the choice of boundary conditions that are imposed on the sphere \cite{Baldo06}.
These shell effects of the order of $\hbar^2/ 2 m R_{\rm cell}^2$ where $R_{\rm cell}$ is the radius of the Wigner-Seitz sphere, may be very large in the deep layers of the crust where the nuclear clusters are very close to each other.
A correct treatment of the ``free'' neutrons requires the application of the band theory originally proposed for describing electrons in solids and recently
applied to neutron star crust \cite{Chamel05,Chamel06}.

\subsection{Band theory}

\begin{figure}
  \includegraphics[height=.3\textheight]{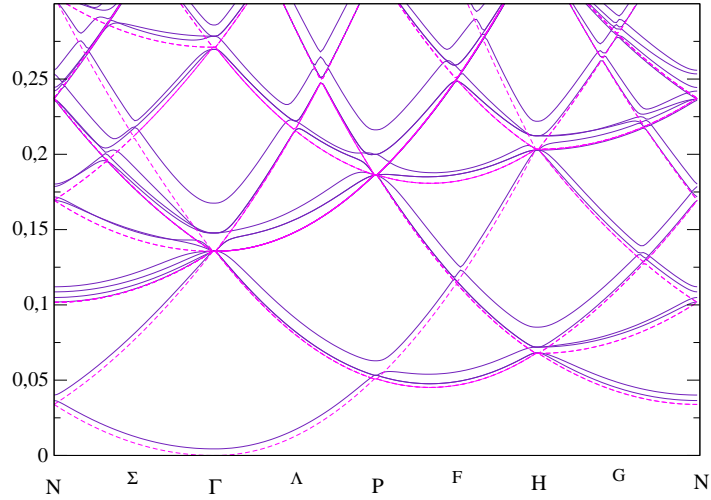}
  \caption{Energy spectrum (in MeV) of unbound neutrons in the outermost layers of the inner crust of neutron stars
  for different symmetry directions in $\mathbf{k}$-space. The dashed line is the energy spectrum of the ideal Fermi gas.
  See reference \cite{Chamel06} for details.}
  \label{fig2}
\end{figure}

The nuclear lattice can be partitioned into identical Wigner-Seitz cells whose shape is determined by the geometry of the lattice. The boundary conditions on the cell
are not arbitrary but are fixed by the Floquet-Bloch theorem. Each single particle quantum state is described by a discrete index $\alpha$ and
by a wave vector $\mathbf{k}$. The energy spectrum is thus formed of a series of sheets or ``bands'' in $\mathbf{k}$-space as illustrated on figure \ref{fig2}.
We have recently applied the band theory by performing 3D calculations with Bloch boundary conditions (see \cite{Chamel06} and references therein), in order to investigate the effects of the nuclear clusters on the dynamical properties of the neutron gas. For this purpose, first ignoring the effects of pairing, we have studied the topology of the Fermi surface by computing its area and comparing it to the area of the corresponding sphere (remembering that the
Fermi volume only depends on the density). We found that at low densities,
meaning that the Fermi wave length of the unbound neutrons is much larger than the lattice spacing, the Fermi surface is nearly spherical,
which is analogous to alkali metals. However at higher densities the Fermi surface is strongly deformed owing to Bragg diffraction as it is also observed for transition
metals. This entails a renormalisation of the neutron mass into an effective mass defined by $m_\star = p_{\rm n}/v_{\rm n}$ and given by a Fermi surface
integral \cite{Carter05}. This effective mass can take very large values, up to about $\sim 15 m_{\rm n}$ at the baryon density $n_b = 0.03$ fm$^{-3}$ \cite{Chamel05}.
We have also shown that in nearly homogeneous neutron star crust matter for which the BCS approximation is valid, the effects of neutron pairing on the effective mass are small and vanish in the limit of a uniform system \cite{Carter05b}. However as pointed out recently by Magierski \cite{Magierski05},
strong inhomogeneities may lead to neutron localisation, which might further enhance the effective neutron mass.

\section{Conclusion}

The evolution of a neutron star is intimately related to the properties of its solid crust. Up to a density of about 
10$^7$ g.cm$^{-3}$, the crust is formed of a body centered cubic crystal of iron $^{56}$Fe, which are fully ionized above $\sim 10^4$ g.cm$^{-3}$. 
With increasing density, the nuclei immersed in a relativistic electron gas, become more and more neutron rich owing to electron capture. At densities around 10$^{11}$ g.cm$^{-3}$ neutrons start to drip out of nuclei and may form Cooper pairs which condense into a superfluid phase. At the bottom of the crust at densities of the order of 10$^{14}$ g.cm$^{-3}$, some calculations predict that nuclei may adopt exotic non spherical shapes, referred as ``pastas''. 

Unlike the nuclei in the shallower layers, the nuclear ``clusters'' in the inner crust cannot be studied in the laboratory due to the presence of ``free'' superfluid neutrons. The structure of the crust has been studied using different approximations and nuclear models. The current state-of-the-art of self-consistent quantum calculations which were pioneered by Negele\&Vautherin\cite{NV73}, is the Hartree-Fock-Bogoliubov approximation. It has been recently shown in this framework that the neutron superfluidity greatly affects the composition of the crust\cite{Baldo05}. 

However much remains to be done. Indeed, in all these calculations the lattice is treated in the Wigner-Seitz approximation in which 
the dripped neutrons are artificially confined in spheres. A much more accurate description of the crust, which is essential in order to interpret observations of neutron stars, should rely on the band theory of solids. We have recently shown that the dynamical properties of the superfluid neutrons are strongly affected by the nuclear clusters by carrying out 3D calculations with Bloch boundary conditions \cite{Chamel06}. This work is a first step towards realistic calculations of the properties of neutron star crust.

\begin{theacknowledgments}
The author gratefully acknowledges financial support from a Marie Curie Intra-European Fellowship of the European Union (contract MEIF-CT-2005-024660).
\end{theacknowledgments}

\bibliographystyle{aipproc}   

\bibliography{toursproc}

\end{document}